\documentclass[aps,twocolumn,superscriptaddress,altaffilletter,lengthcheck,
tightenlines,showpacs,showkeys]{revtex4}

\newcommand{\ben}{\begin{eqnarray}}
\newcommand{\een}{\end{eqnarray}}
\newcommand{\be}{\begin{equation}}
\newcommand{\ee}{\end{equation}}
\newcommand{\ba}{\begin{eqnarray}}
\newcommand{\ea}{\end{eqnarray}}
\newcommand{\n}{\label}

\newcommand{\ga}{\gamma}
\newcommand{\ro}{\rho}

\usepackage{graphicx}

\usepackage{epstopdf}

\usepackage[dvipdf]{epsfig}
\usepackage{color}
\usepackage{multirow}

\begin{document}
\title{Dark matter interacts with variable vacuum energy}

\author{Iv\'an E. S\'anchez G.}\email{isg.cos@gmail.com}
\affiliation{Departamento de Matem\'{a}tica, Facultad de Ciencias Exactas y Naturales, Universidad de Buenos Aires and IMAS, CONICET, Ciudad Universitaria, Pabell\'on I, 1428 Buenos Aires, Argentina}

\date{\today}
\bibliographystyle{plain}

\begin{abstract}
We investigate a spatially flat Friedmann-Robertson-Walker (FRW) scenario with two interacting components, dark matter and variable vacuum energy (VVE) densities, plus two decoupled components, one is a baryon term while the other behaves as a radiation component. We consider a linear interaction in the derivative dark component density. We apply the $\chi^2$ method to the observational Hubble data for constraining the cosmological parameters and analyze the amount of dark energy in the radiation era for the model. It turns out that our model fulfills the severe bound of $\Omega_{x}(z\simeq 1100)<0.009$ at $2\sigma$ level, so is consistent with the recent analysis that include cosmic microwave background anisotropy measurements from Planck survey, the future constraints achievable by Euclid and CMBPol experiments, reported for the behavior of the dark energy at early times, and fulfills the stringent bound $\Omega_{x}(z\simeq 10^{10})<0.04$ at $2\sigma$ level in the big-bang nucleosynthesis epoch. We also examine the cosmic age problem at high redshift associated with the old quasar APM 08279+5255 and estimate the age of the universe today.


\end{abstract}
\vskip 1cm



\date{\today}
\bibliographystyle{plain}

\maketitle

\section{Introduction}
Around the last fifteen years, studies of the available high quality cosmological data, the brightness of a class of supernovas (SNIa) \cite{Riess1} \cite{Riess2} \cite{Perlm} \cite{Astier}, the spectra of the cosmic microwave background (CMB) anisotropies \cite{Spergel} \cite{Spergel1}, the baryon acoustic oscillations (BAO) in the Sloan digital sky survey (SDSS) luminous galaxy sample \cite{Adelman} \cite{Tegmark} \cite{Tegmark1}, have converged towards a cosmic expansion history that involves a spatially flat geometry and a recent accelerating period of the Universe. This faster expansion phase has been attributed to the misterious dark energy component with negative pressure, which represents more than the $70\%$ of the total energy of the Universe. Despite the arrival of several observational data of ever increasing quality and quantity, the insight into the fundamental nature of the dark energy component is still unknown, owing to the effects observable through its gravitational interaction at cosmological distances. The simplest type of dark energy corresponds to a positive cosmological constant $\Lambda$.

Evidence indicates that in the cosmological history of the Universe, the matter and space-time emerged from singularity and evolved through four different eras: early inflation, radiation, dark matter and dark energy dominated eras. During the radiation and dark matter stages, the expansion slows down while in the inflation and dark energy eras it speeds up. Besides, the necessity of a dark matter component comes from astrophysical evidences of colliding galaxies, a power spectrum of clustered matter or gravitational lensing of mass distribution \cite{Drees} \cite{Garrett}. Today, astrophysical observations suggest that dark matter is a substantial component to the Universe's total matter density and represents nearly $23\%$ of the total energy matter of the Universe. This nonbaryonic invisible component is the major agent responsible for the large-structure formation in the Universe.

Considering the mechanisms that govern the nature of both dark components, we could propose the existence of an exchange of energy between them, i.e., assume that the dark matter feels the presence of dark energy through a gravitational expansion of the Universe but also that they can interact with each other \cite{Luis} \cite{LuisM8}. A coupling between dark matter and dark energy changes the background evolution of the dark sector, giving rise to a rich cosmological dynamics compared with non interacting models.

Within the framework of interaction models, the source of Einstein equations which describes the dynamics of the Universe at large scale includes an aggregate of different material fluids that are conserved individually or with interactions \cite{Chen}, being this the simplest hypotesis to start with. Following the observational evidences we will consider four fundamental components: radiation, baryons, dark matter and dark energy.

Confronting these models with the observational data could lead to new insights about the properties of dark matter and dark energy. For example, the fraction of dark energy at recombination era should fulfill the bound $\Omega_{x}(z\simeq 1100)<0.1$ in order to the dark energy model be consistent with the big-bang nucleosynthesis (BBN) data. Unraveling the nature of dark as well as their properties to high redshift could give an invaluable guide to the physics behind the recent acceleration of the universe \cite{Luis4}-\cite{Luis7}. The current constraints on the amount of dark energy at early times come from the Planck mission, the cosmological data analyzed has led to an upper bound of $\Omega_{x}(z\simeq 1100)<0.009$ with $95\%$ confidence level (CL) \cite{Planck2013}. Besides, future surveys such as Euclid or CMBPol will be able to constrain on the fraction of early dark energy. The joint analysis based on Euclid+CMBPol data leads to $\Omega_{x}(z\simeq 1100) <0.00092$ while the joint analysis of Euclid+Planck data will be less restrictive yielding $\Omega_{\rm x}(z\simeq 1100) <0.0022$ \cite{HollEuc}.

The aim of this article is to examine the exchange of energy between dark matter and the dark energy in a model with a linear interaction in the derivative of the energy density, and two decoupled components. We constrain the cosmic set of parameters by using the updated Hubble data and the severe bounds reported by Planck mission on early dark energy.

\section{The Model}
We consider a spatially flat isotropic and homogeneous universe described by Friedmann-Robertson-Walker (FRW) spacetime. The universe is filled with four components, one very close to radiation, baryonic matter, dark matter and variable vacuum energy (VVE), the last two of them interacts and the first are decoupled components. The evolution of the FRW universe is governed by the Friedmann and conservation equations,
\be
\label{E1a}
3H^{2}=\ro_T=\rho_{r}+\rho_{b}+\ro_{m}+\ro_{x},
\ee
\be
\label{E1c}
\dot\ro_{r}+3H\ga_{r}\ro_{r}=0,~~~\dot\ro_{b}+3H\ga_{b}\ro_{b}=0,
\ee
\be
\label{E1b}
\dot{\ro}_{m}+\dot{\ro}_{x}+3H(\ga_m\rho_{m}+\ga_x\rho_{x})=0,
\ee
where $H=\dot a/a$ is the Hubble expansion rate and $a(t)$ is the scale factor. The equation of state for each species, with energy densities $\ro_{\rm i}$, and pressures $p_{\rm i}$, take a barotropic form $p_{\rm i}=(\gamma_{\rm i}-1)\ro_{\rm i}$, then the constants $\ga_{\rm i}$ indicate the barotropic index of each component being ${\rm i}=\{x,m,b,r\}$, so  that  $\gamma_{x}=0$, $\ga_{b}=1$, whereas $\gamma_{r}$ and $\ga_{m}$ will be estimated later on. So, $\rho_{x}$ plays the role of a decaying vacuum energy or variable cosmological constant, $\ro_{b}$ represents a pressureless barionic matter,  $\ro_{r}$ is close to a radiation component and $\rho_{m}$ can be associated with dark matter.

Solving the linear system of equation (\ref{E1b}) along with $\ro=\ro_{m}+\ro_{x}$ we acquire both dark densities as functions of $\ro$ and $\ro'$
\be
\n{04}
\ro_{m}= - \frac{\ga_{x} \ro +\ro '}{ \ga_{m}-\ga_{x}}, \qquad \ro_x= \frac{\ga_{m} \ro +\ro '}{ \ga_{m}-\ga_{x}},
\ee
where we have used the variable $\eta=ln(a/a_0)^3$ with $a_0$ the present value of the scale factor ($a_0=1$). We suppose that there is no interaction between the radiation, baryons and the dark sector, so the energy density is conserved and the prime indicates differentiation with respect to the new time variable $'\equiv d/d\eta$. Under this situation, Eqs. (\ref{E1c}) leads to the energy densities for radiation and baryonic matter, $\ro_{r}\sim a^{-3\ga_r}$ and $\ro_{b}\sim a^{-3}$, respectively.

In order to continue the analysis of the interacting dark sector, we introduce an energy transfer between the two fluids, by separating the conservation equation like
\be
\label{rod3}
{\ro'}_{m}+\ga_m\rho_{m}=-Q,  \qquad {\ro'}_{x}+\ga_x\rho_{x}=Q.
\ee
We have consider a coupling with a factorized $H$ dependence in the form $3HQ$, where $Q$ indicates the energy exchange between the two dark components. From Eqs. (\ref{04}) and (\ref{rod3}), we obtain the source equation \cite{Luis} for the energy density $\ro$ of the dark sector
\be
\n{rod4}
\ro''+(\ga_{m}+ \ga_{x})\ro' + \ga_{m}\ga_{x}\ro =  Q(\ga_{m}-\ga_{x}).
\ee
Here, the interaction $Q$ between both dark components takes the form, $Q=\alpha\ro'$, being $\alpha$ the coupling constant that measures the strength of the interaction in the dark sector. This kind of interaction is now analyzed under the view of the new observations and gives rise to a dark energy model that can be viewed as a running cosmological constant or a decaying vacuum energy \cite{Basilakos}, \cite{Lima}. The aim of this work is to explore this model characterized by a variable cosmological constant with the observational constraints coming from the behavior of dark energy at early times.

Replacing the specific form of $Q$ into the source equation (\ref{rod4}) it turns into a second order differential equation for the total energy density $\ro$. Inserting $\ga_{x}=0$ into the latter equation one gets a first order linear differential equation
\be
\label{rod5}
\ro'=\ga_{m}[\left(\alpha-1\right)\ro + {\cal C}],
\ee
where $ {\cal C}$ is an integration constant. From the latter, we can see that replacing Eq. (\ref{rod5}) in the second equation of (\ref{04}), the dark energy of this model can be considered as a VVE,
\be
\label{L}
\ro_{x}=\alpha\ro + {\cal C}=\Lambda,
\ee
In the future when the dark energy dominates the whole dynamics of the universe, it will be $\ro_{T}\approx\ro_{x}$, so the equation (\ref{L}) goes to the well-known $\Lambda(H)$ model, $\Lambda\simeq\alpha H^{2} + {\cal C}$, \cite{Basilakos}, \cite{Lima}.

In order to get $\rho(a)$ we need to express the first-order linear differential equation (\ref{rod5}) as an integration by quadrature as follows,
\be
\label{rod12}
\ro=\frac{{\cal K}}{1-\alpha}(1+z)^{3(1-\alpha)\ga_m}+\frac{{\cal C}}{1-\alpha},
\ee
where ${\cal K}$ is an integration constant. We write $\ro$ in terms of the redshift $z$, considering the relation between the scalar factor and the redshift, $z+1=1/a$. Using  the present-density parameters $\Omega_{i0}=\ro_{i0}/3H_{0}^2$ along with the flatness condition, $1=\Omega_{r0}+\Omega_{b0}+\Omega_{x0}+\Omega_{m0}$, we can write the integration constants ${\cal K}$ and ${\cal C}$ in terms of the observational density parameters:
\be
\label{K1}
{\cal K}=3{H_0}^2\Omega_{m0},
\ee
\be
\label{C1}
{\cal C}=3{H_0}^2[\Omega_{x0}-\alpha(\Omega_{x0}+\Omega_{m0})].
\ee

In this case, the total energy density $\ro_{T}/3H_{0}^{2}$ is given by
\[\frac{\ro_{T}}{3H_{0}^2}=(1-\Omega_{b0}-\Omega_{x0}-\Omega_{m0})(1+z)^{3\ga_r}+\Omega_{b0}(1+z)^{3}\]
\be
\label{H2}
+\frac{\Omega_{m0}}{1-\alpha}(1+z)^{3(1-\alpha)\ga_m}+\Omega_{x0}+ (1-\frac{1}{1-\alpha})\Omega_{m0}.
\ee
So, the model has seven independent parameters ($H_0$,$\Omega_{b0}$,$\Omega_{x0}$,$\Omega_{m0}$,$\alpha$,$\ga_{r}$,$\ga_{m}$) to be completely specified. From (\ref{H2}) we see that the Universe is dominated by radiation at early times where the dark components are negligible. After this epoch pressureless baryonic matter dominates followed by an era governed by dark matter when $(1-\alpha)\gamma_{m}\simeq1$. Finally, the universe exhibits a de Sitter phase at late times.  Then, the interaction allows a smooth transition between a dark matter era in the distant past (intermediate regime) to a speeding up stage at late times. For the limit cases, when $z\rightarrow-1$ the energy densities goes to $\ro_{m}\rightarrow 0$ and $\ro_{x}\rightarrow \frac{C}{1-\alpha}$, and when $z\rightarrow\infty$, $\ro_{m}\rightarrow\infty$ and $\ro_{x}\rightarrow\infty$, if $\alpha <1$; such can be verified by using the best-fit values of the cosmological parameters found in the next section.

\section{Observational Hubble data constraints}
We are going to find a qualitative estimation of the cosmological parameters for the model, with an interaction in the dark sector given by $Q=\alpha\ro'$, plus the decoupled radiation and baryonic components, using the observational $H(z)$ data \cite{Stern} \cite{Verde} \cite{Moresco} \cite{Busca} \cite{Zhang} \cite{Blake} \cite{Chuang}. The values of the function $H(z)$ are directly obtained from the cosmological observations, so this function plays a fundamental role in understanding the properties of the dark sector. From the relation $dt/dz=-(1+z)H(z)$ \cite{Stern} \cite{Verde} a measurement of the differential ages $dz/dt$ at different redshifts, gives the Hubble function. The statistical method requires the compilation of the observed $H_{obs}$ \cite{Chuang} and the best value for the present time $z=0$, adjusted according to \cite{Ries}. The bibliography \cite{W-7}, \cite{WMAP9}, \cite{Ratra} shows $H_{obs}$ for different redshifts with the corresponding 1$\sigma$ uncertainties. The probability distribution for the $\theta$-parameters is $P(\theta)=\aleph\exp^{-\chi^2(\theta)/2}$ (see e.g. \cite{Press}) being $\aleph$ a normalization constant. The parameters of the model are determined by minimizing
\be
\label{chi}
\chi^2(\theta)=\sum^{N=29}_{i=1}\frac{[H(\theta;z_i)-H_{obs}(z_i)]^2}{\sigma^2(z_i)},
\ee
where $H_{obs}(z_i)$ is the observed value of $H(z)$ at the redshift $z_i$, $\sigma(z_i)$ is the corresponding 1$\sigma$ uncertainty and $H(\theta,z_i)$ is the Hubble function (\ref{H2}) evaluated at $z_i$. The $\chi^2$ function reaches its minimum value at the best fit value $\theta_c$ and the fit is good when $\chi^{2}_{min}(\theta_c)/(N-n)\leq1$ where $n$ is the number of parameters \cite{Press}. Here, $N=29$ is the number of the data and $n=2$. The variable $\chi^2$ is a random variable that depends on $N$ and its probability distribution is a $\chi^2$ distribution for $N-n$ degrees of freedom.
In our case, we consider $\theta=(H_0,\Omega_{b0},\Omega_{x0},\Omega_{m0},\alpha,\gamma_r,\gamma_m)$ plus the constraint on the density parameters to assure the flatness condition $(\Omega_{r0}=1-\Omega_{b0}-\Omega_{x0}-\Omega_{m0})$; so we have seven independent parameters. For a given pair $(\theta_1,\theta_2)$ of independent parameters, fixing the other ones, we will perform the statistical analysis by minimizing the $\chi^2$ function to obtain the best-fit values of the random variables. The confidence levels $1\sigma$ $(68.3\%)$ or $2\sigma$ $(95.4\%)$ will satisfy $\chi^2(\theta)-\chi^{2}_{min}(\theta_c)\leq2.30$ or $\chi^2(\theta)-\chi^{2}_{min}(\theta_c)\leq6.17$ respectively.

\begin{figure}[hbt!]
\begin{minipage}{1\linewidth}
\resizebox{1.6in}{!}{\includegraphics{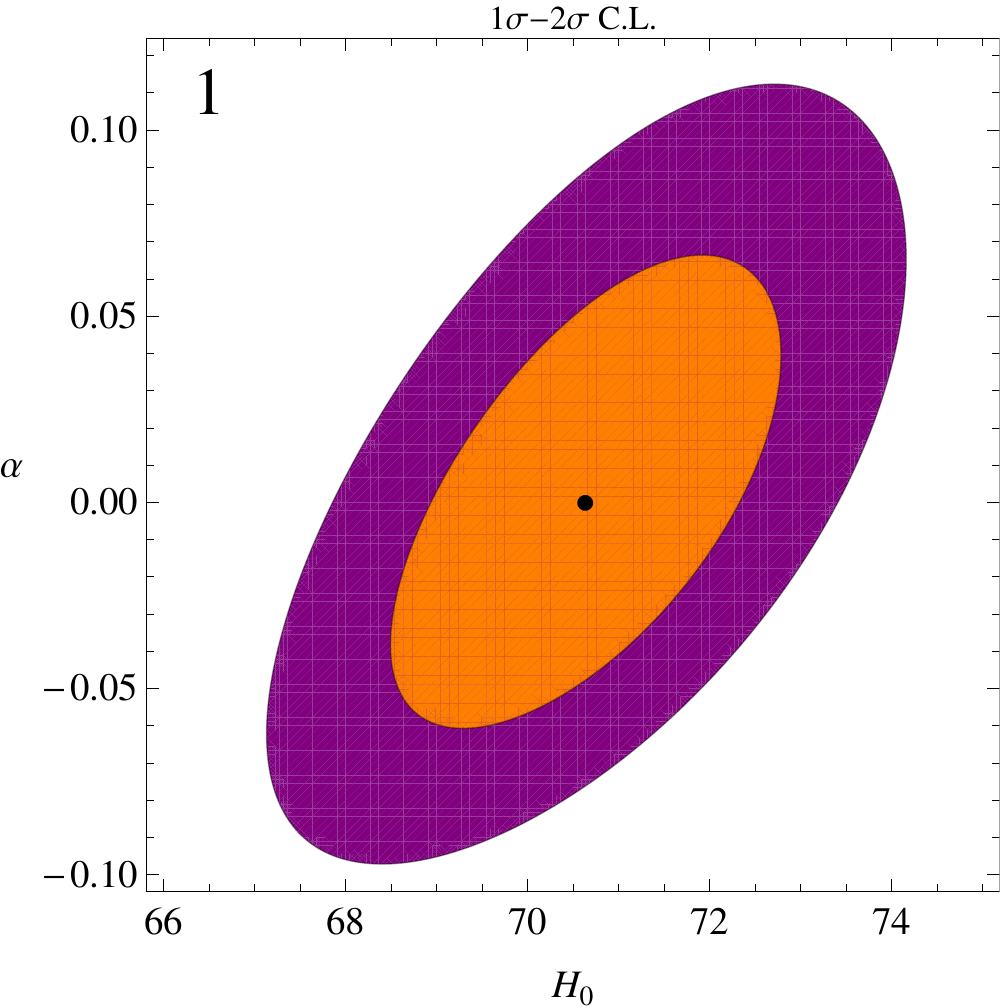}}\hskip0.06cm
\resizebox{1.6in}{!}{\includegraphics{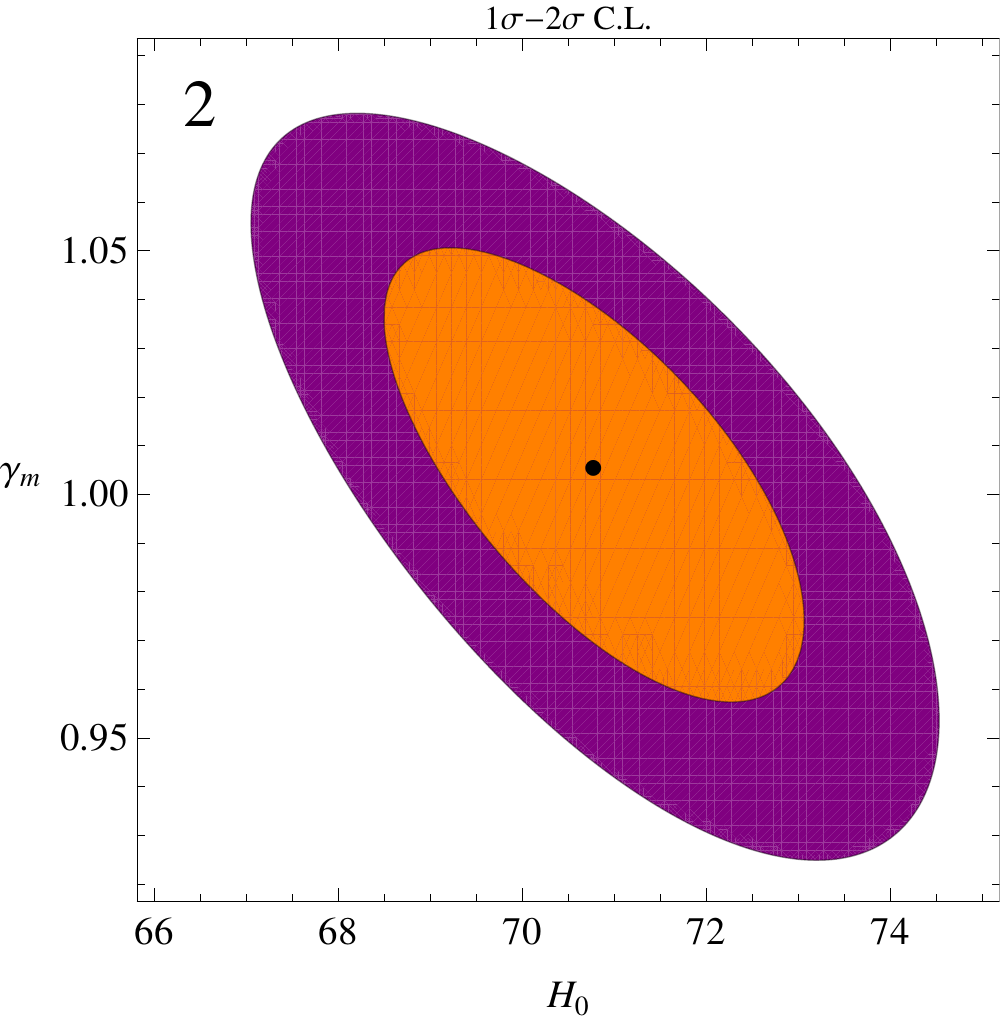}}\hskip0.06cm
\resizebox{1.6in}{!}{\includegraphics{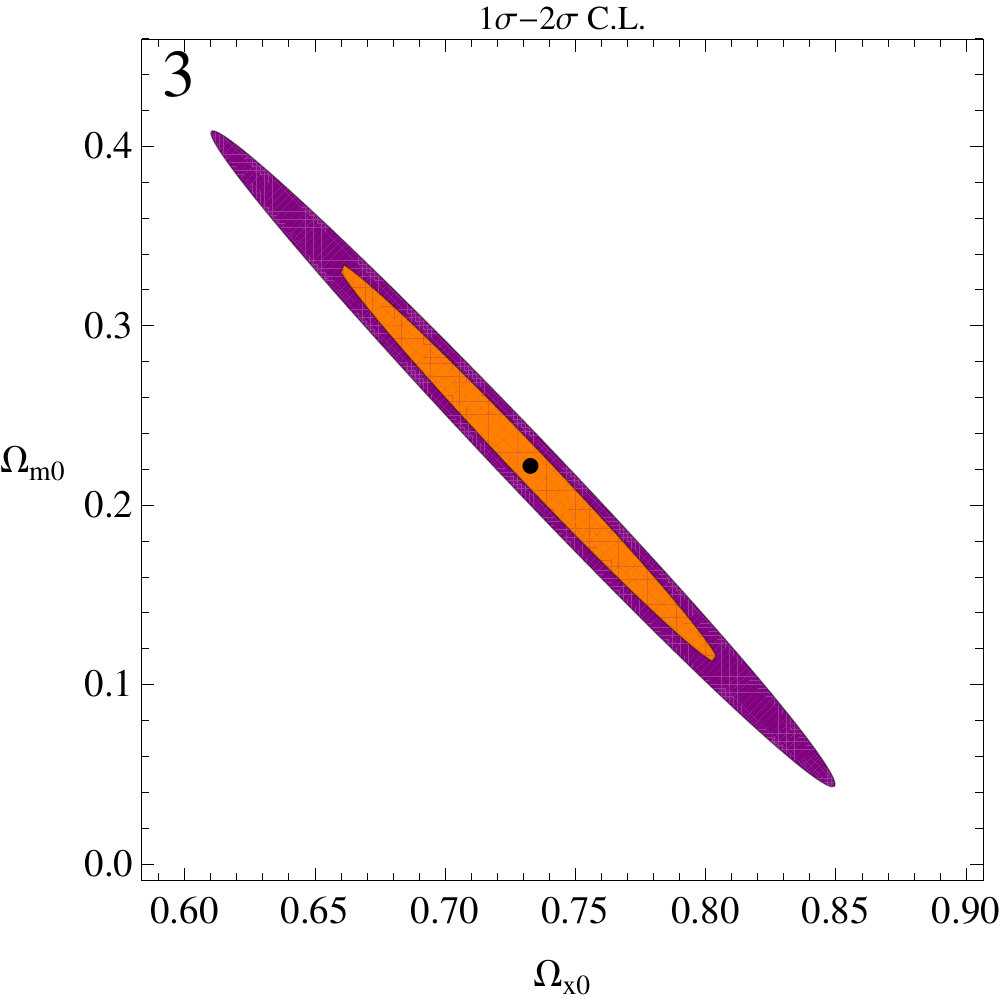}}\hskip0.06cm
\resizebox{1.6in}{!}{\includegraphics{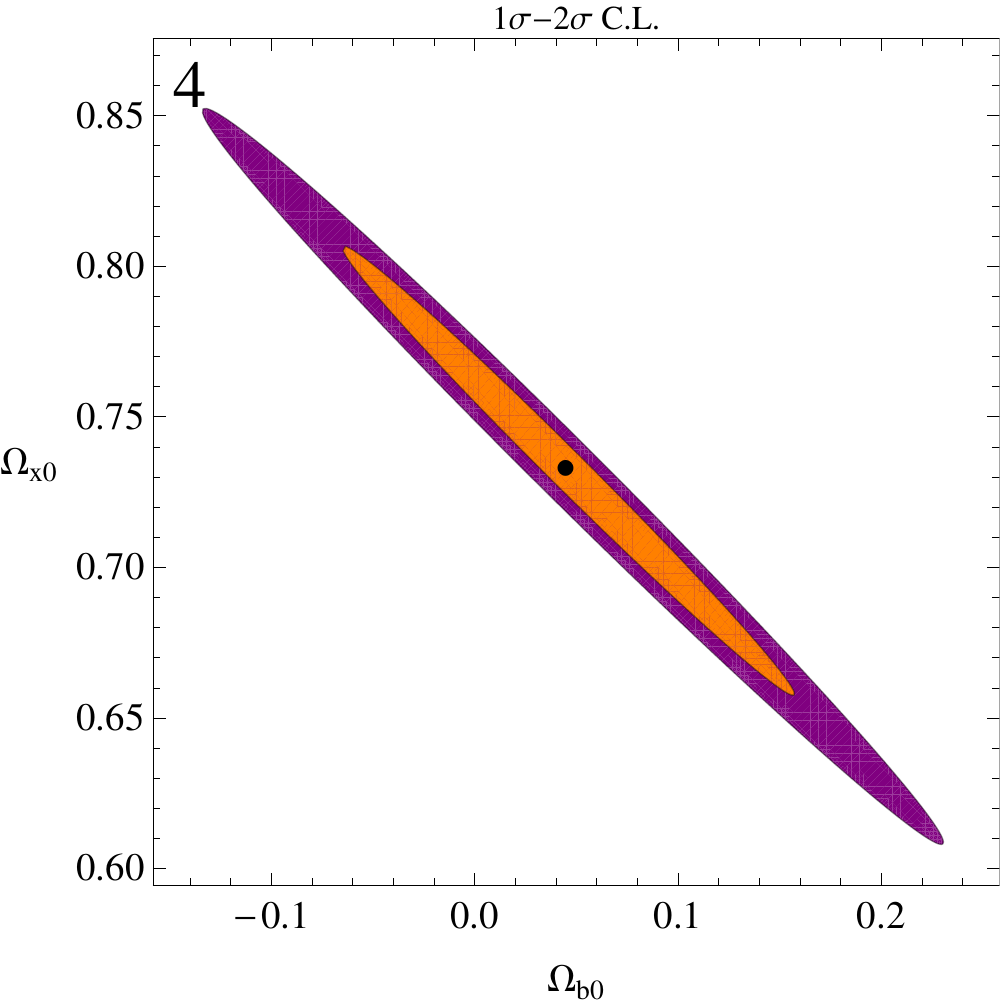}}\hskip0.06cm
\resizebox{1.6in}{!}{\includegraphics{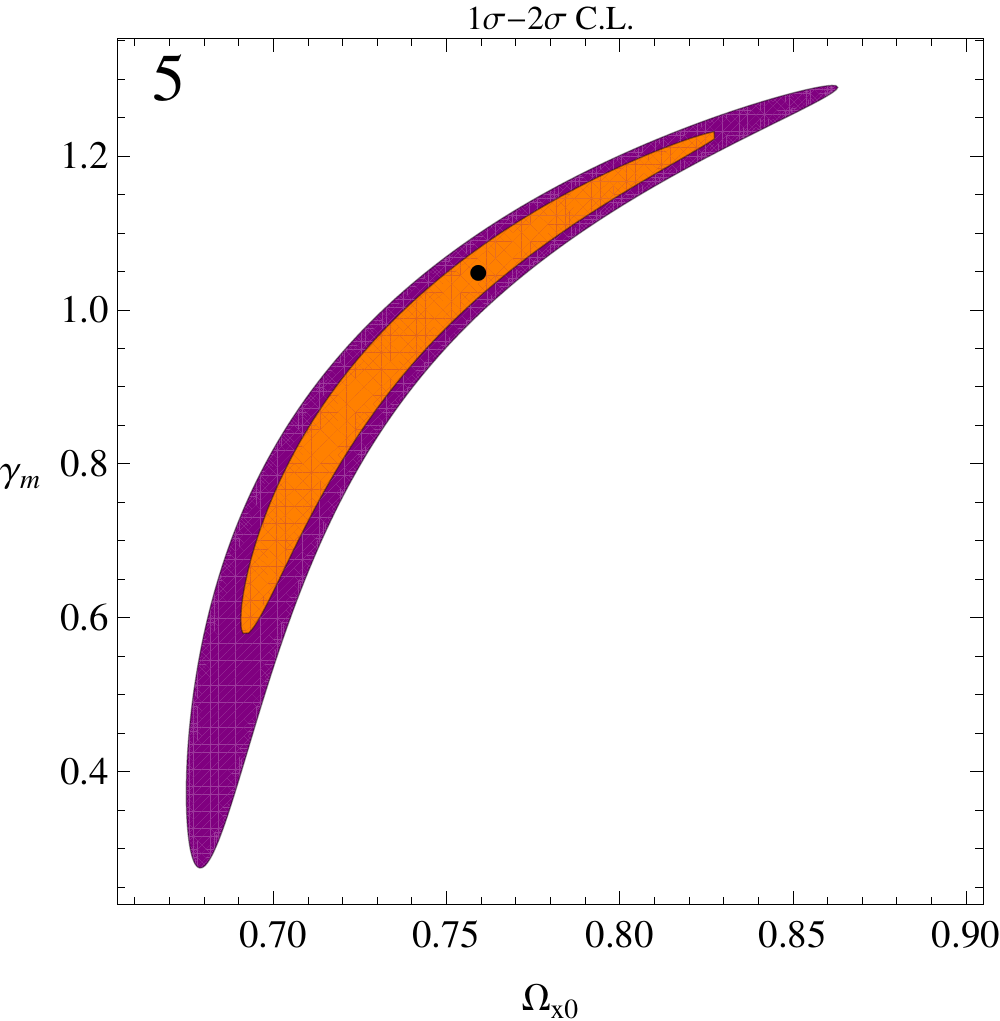}}\hskip0.06cm
\resizebox{1.6in}{!}{\includegraphics{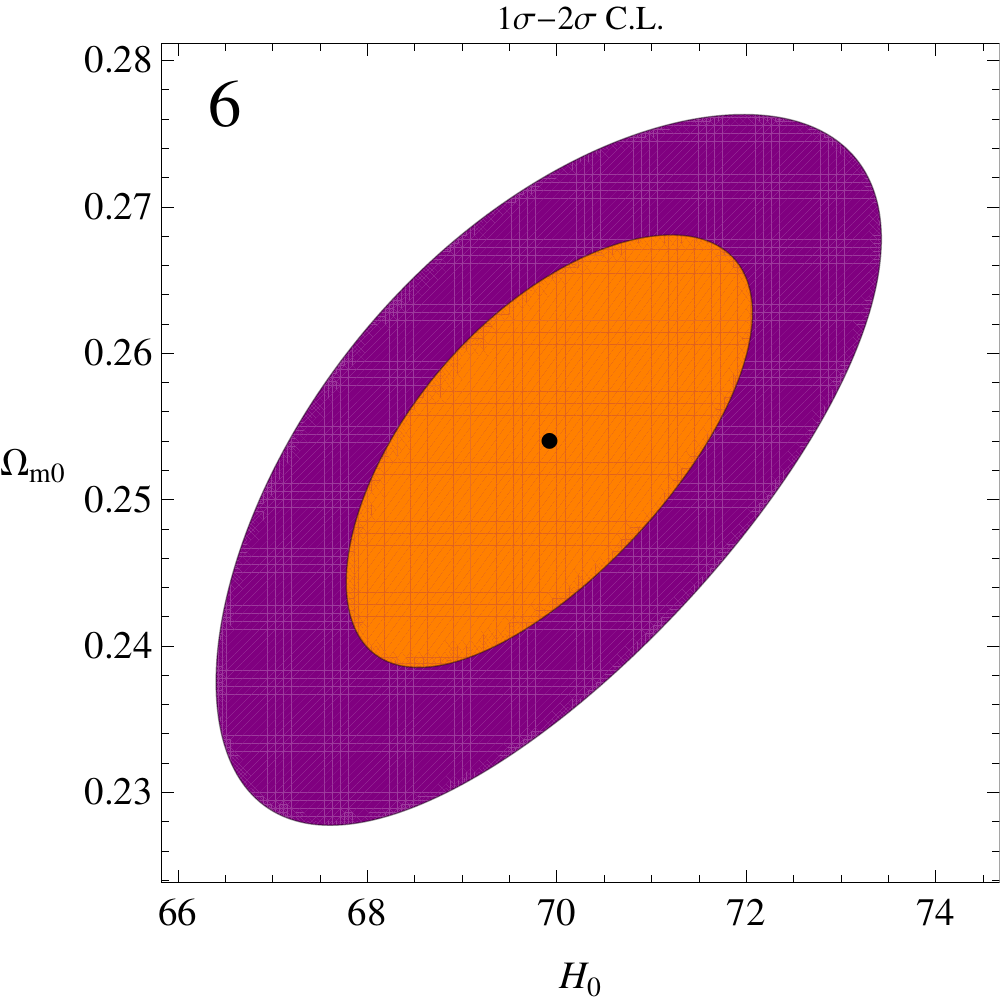}}\hskip0.06cm
\caption{\scriptsize{Two-dimensional C.L. associated with $1\sigma$, $2\sigma$ for different $\theta$ planes with the interaction $Q=\alpha\ro'$.}}
\label{F11}
\end{minipage}
\end{figure}

\begin{figure}[hbt!]
\begin{minipage}{1\linewidth}
\resizebox{1.69in}{!}{\includegraphics{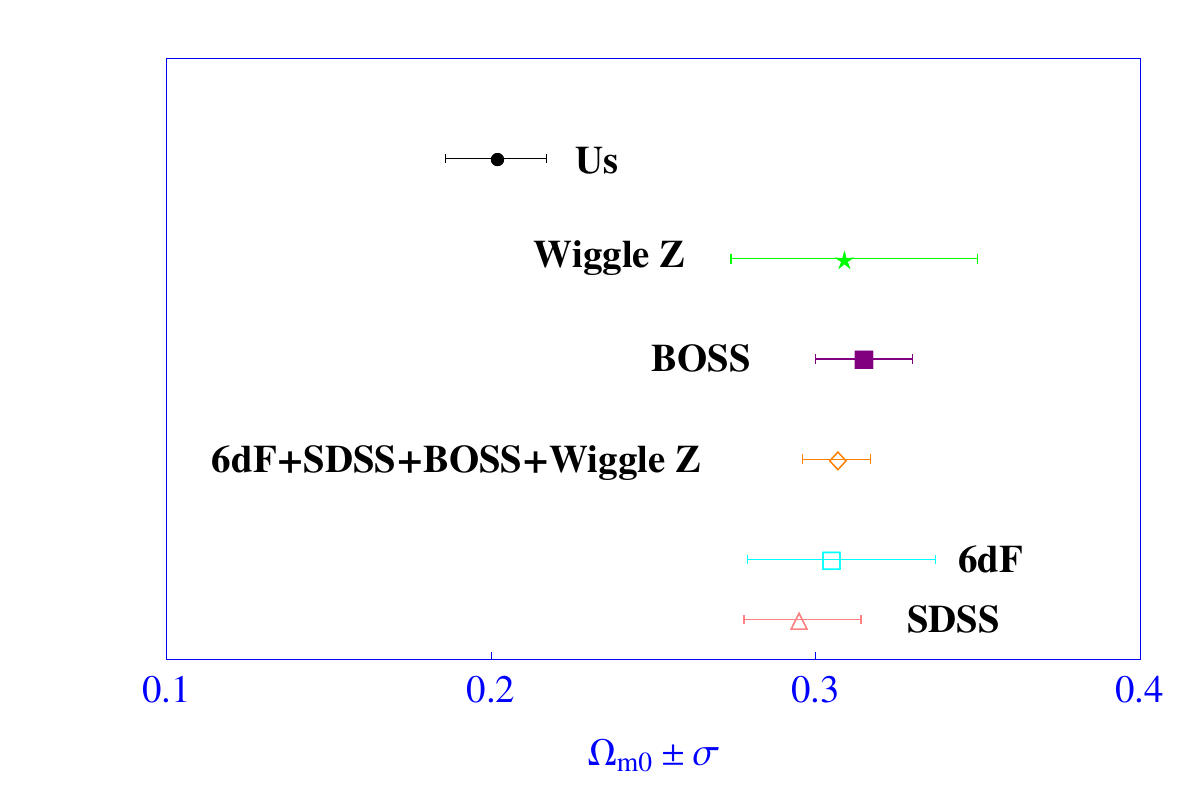}}\hskip0.06cm
\resizebox{1.69in}{!}{\includegraphics{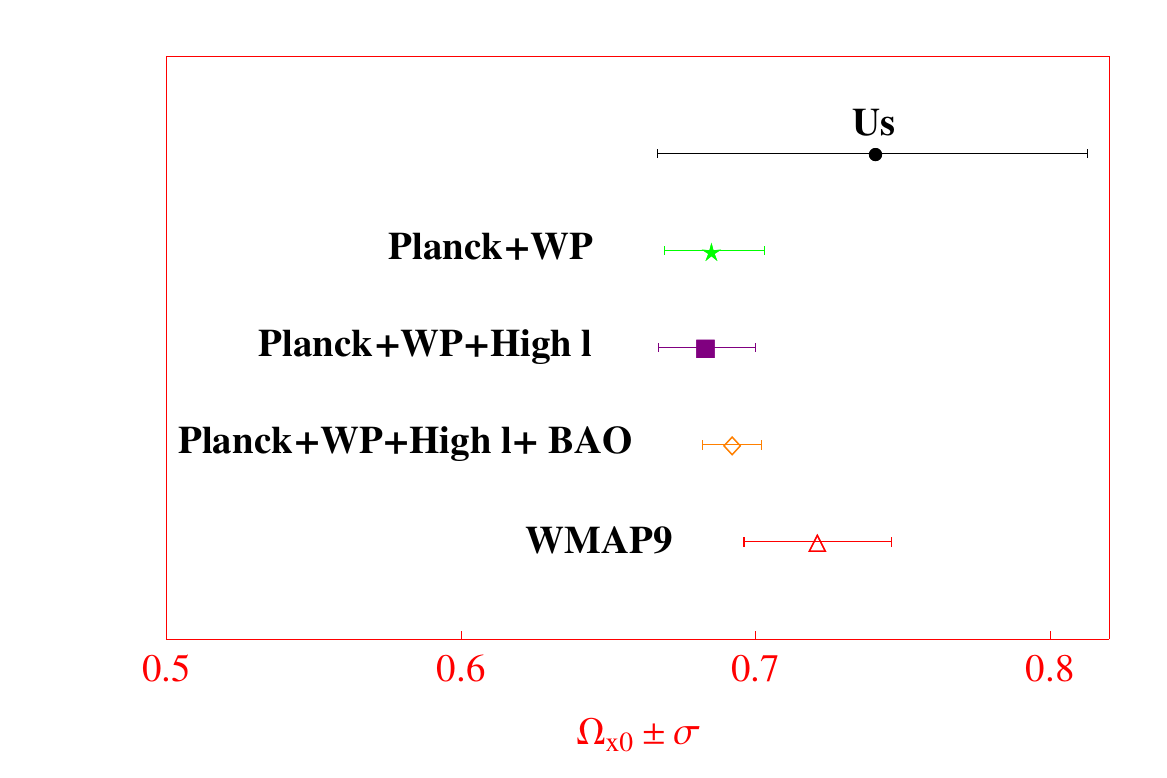}}\hskip0.06cm
\caption{\scriptsize{Comparison of dark matter and dark energy amounts, with estimates of $\sigma$ errors, from a number of different methods.}}
\label{FO}
\end{minipage}
\end{figure}

The confidence levels (C.L) obtained with the standard $\chi^2$ function for two independent parameters are shown in Fig. (\ref{F11}), whereas the estimation is briefly summarized in Table (\ref{I1}); reporting their corresponding marginal $1\sigma$ error bars \cite{Sivia}. At number $1$ of the table (\ref{I1}), we find the best fit at $(H_0,\alpha)=(70.45^{+2.13}_{-2.16}{\rm km s^{-1} Mpc^{-1}}$, $0.00007^{+0.06604}_{-0.06000})$ with $\chi^{2}_{d.o.f}=0.709$, that fulfills the goodness condition $\chi^{2}_{d.o.f}<1$. At $3$, we get the best fit at the independence parameters $(\Omega_{x0},\Omega_{m0})=(0.733^{+0.068}_{-0.072}$, $0.222^{+0.108}_{-0.107})$ with $\chi^{2}_{d.o.f}=0.758$ by using the priors ($H_{0}$, $\Omega_{b}$, $\alpha$, $\ga_{r}$, $\ga_{m}$)$=$($69$, $0.0449$, $0.0001$, $4/3$, $1.014$); so the values obtained for  $\Omega_{x0}$ and  $\Omega_{m0}$ are in agreement with the data released by Planck Mission \cite{Planck2013} or with the data coming from  the WMAP-9 project \cite{WMAP9} [see Fig. (\ref{FO})]. Indeed, Planck+WMAP data indicate that the vacuum energy amount is $0.685 ^{+0.018}_{-0.016}$ at $68\%$ C.L,  Planck+WMAP+high L data lead to  $0.6830 ^{+0.017}_{-0.016}$ at $68\%$ C.L  whereas the joint statistical analysis on  Planck+WMAP+high L+ BAO gives   $0.692\pm 0.010$ at $1\sigma$ level \cite{Planck2013}. The approximate constraints on the present day value of dark matter with $68\%$ errors show that Wiggle-Z data give $\Omega_{m0}=0.309^{+0.041}_{-0.035}$, while Boss experiment seems to increase the dark matter amount in $0.019\%$, thus $\Omega_{m0}=0.315^{+0.015}_{-0.015}$; whereas the joint statistical analysis with the data 6dF+SDSS+BOSS+Wiggle-Z leads to $\Omega_{m0}=0.307^{+0.010}_{-0.011}$ at $68\%$ confidence level \cite{Planck2013}, so our result for $\Omega_{m0}$ overlaps with the observational data [see Fig. (\ref{FO})]. Besides, we find the best fit at $(\Omega_{x0}, \ga_{m})=( 0.759^{+0.066}_{-0.067}, 1.049^{+0.181}_{-0.460})$ with $\chi^2_{\rm d.o.f}=0.762$, pointing that the dark matter is not pressureless provided the barotropic index is slightly greater than the unity [cf. Table (\ref{I1})]. Also, the statistical analysis leads to $(\Omega_{b0}, \Omega_{x0})=( 0.045^{+0.107}_{-0.104}, 0.733^{+0.070}_{-0.074}$ with $\chi^2_{\rm d.o.f}= 0.758$, that is in agreement with the $\Omega_{b0}=0.035\pm 0.001$ coming from the WMAP-9 \cite{WMAP9} and the $\Omega_{b0}=0.034\pm 0.001$ at $68\%$ C.L of the Planck Mission \cite{Planck2013}. Regarding the present day value of the Hubble parameter, we find that varies over a wide range, $H_{0} \in [69.92^{+2.11}_{-2.15}; 70.77^{+2.29}_{-2.27} ]{\rm km~s^{-1}\,Mpc^{-1}}$. From the Planck+WMAP+high L analysis, it was found that $H_{0} = (67.3\pm 1.2) {\rm km s^{-1} Mpc^{-1}}$ at $68\%$ C.L. \cite{Planck2013}. A low value of $H_{0}$ has been found in other CMB experiments, most notably from the recent WMAP-9 analysis. Fitting the base $\Lambda$CDM model for the WMAP-9 data, it was found $ H_{0} =(70.0 \pm  2.2) {\rm km s^{-1} Mpc^{-1}} $ at $68\%$ C.L. \cite{WMAP9}. Then, one of our best estimations $ H_{0} =69.92^{+2.11}_{-2.15} {\rm km s^{-1} Mpc^{-1}} $ at $68\%$ C.L  is perfectly in agreement with the values reported by WMAP-9 and the Planck+WMAP+high L data.

\vskip 0.1cm
\begin{table}[ht!]
\centering
\scalebox{0.57}{
\begin{tabular}{|l|l|l|l|}
  \hline
  \multicolumn{4}{|c|}{${\rm 2D}$ ${\rm Confidence}$ ${\rm level}$ ${\rm for}$ ${\rm Q=\alpha\ro'}$} \\
  \hline
  ${\rm N^{o}}$ & ${\rm Priors}$ & ${\rm Best}$ ${\rm fits}$ & ${\rm {\chi_{d.o.f}^2}}$ \\
  \hline
  1 & ($\Omega_{b}$, $\Omega_{x}$, $\Omega_{m}$, $\ga_{r}$, $\ga_{m}$)=(0.049, 0.741, 0.209, 4/3, 1.005) & ($H_{0}$, $\alpha$)=($70.45^{+2.13}_{-2.16}$, $0.00007^{+0.06604}_{-0.06000}$) & 0.709 \\
  \hline
  2 & ($\Omega_{b}$, $\Omega_{x}$, $\Omega_{m}$, $\alpha$, $\ga_{r}$)=(0.044, 0.744, 0.212, 0.00001, 4/3) & ($H_{0}$, $\ga_{m}$)=($70.77^{+2.29}_{-2.27}$, $1.005^{+0.045}_{-0.048}$) & 0.707 \\
  \hline
  3 & ($H_{0}$, $\Omega_{b}$, $\alpha$, $\ga_{r}$, $\ga_{m}$)=(69, 0.0449, 0.0001, 4/3, 1.014) & ($\Omega_{x}$, $\Omega_{m}$)=($0.733^{+0.068}_{-0.072}$, $0.222^{+0.108}_{-0.107}$) & 0.758 \\
  \hline
  4 & ($H_{0}$, $\Omega_{m}$, $\alpha$, $\ga_{r}$, $\ga_{m}$)=(69, 0.222, 0.00001, 4/3, 1.014) & ($\Omega_{b}$, $\Omega_{x}$)=($0.045^{+0.107}_{-0.104}$, $0.733^{+0.070}_{-0.074}$) & 0.758 \\
  \hline
  5 & ($H_{0}$, $\Omega_{b}$, $\Omega_{m}$, $\alpha$, $\ga_{r}$)=(69, 0.031, 0.210, 0.00001, 1.325) & ($\Omega_{x}$, $\ga_{m}$)=($0.759^{+0.066}_{-0.067}$, $1.049^{+0.181}_{-0.460}$) & 0.762 \\
  \hline
  6 & ($\Omega_{b}$, $\Omega_{x}$, $\alpha$, $\ga_{r}$, $\ga_{m}$)=(0.053, 0.692, 0.00001, 4/3, 0.95) & ($H_{0}$, $\Omega_{m}$)=($69.92^{+2.11}_{-2.15}$, $0.254^{+0.014}_{-0.015}$) & 0.757 \\
  \hline
\end{tabular}}
\caption{\label{I1} \scriptsize{We show the observational bounds for the 2-D C.L. obtained in Fig. (1) by varying two cosmological parameters.}}
\end{table}

\begin{figure}[hbt]
\begin{center}
\includegraphics[width=8cm]{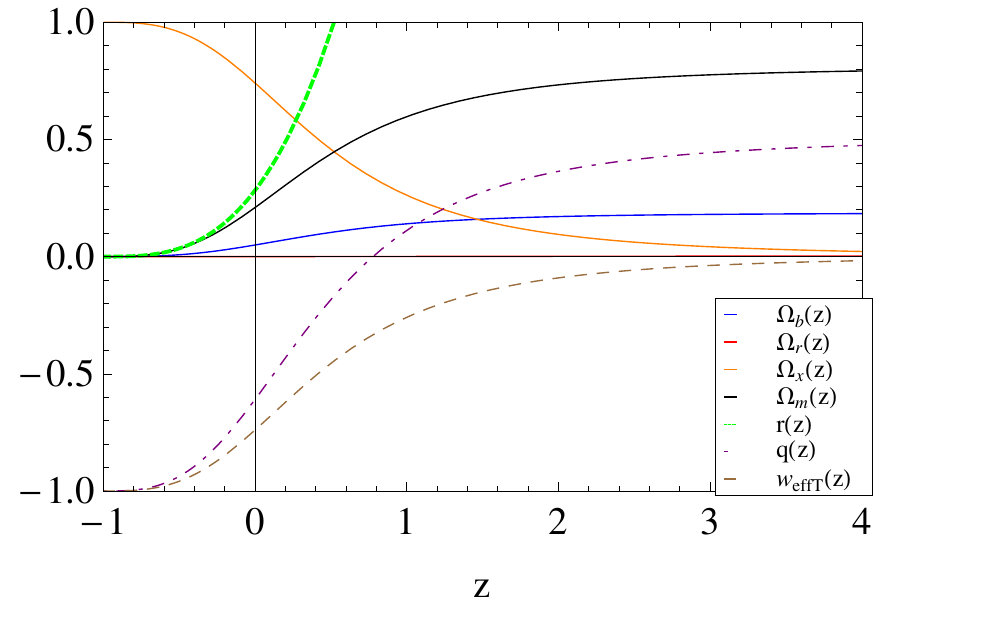}
\caption{\scriptsize{Plot of $\Omega_{b}(z)$, $\Omega_{r}(z)$, $\Omega_{x}(z)$, $\Omega_{m}(z)$, $r(z)$, $q(z)$ and $\omega_{effT}(z)$, using the best-fit values obtained with the Hubble data for different $\theta$ planes.}}
\label{F22}
\end{center}
\end{figure}

There are other cosmological relevant parameters [see Table (\ref{II2})], such as the deceleration parameter at the present time $q(z=0)=q_0$. The figure (\ref{F22}) shows the behavior of a deceleration parameter, the density parameters and the equation of states with the redshift. In particular, present-day value of $q(z=0) \in [-0.62; -0.56]$ as stated in the  WMAP-9 report \cite{WMAP9}. The total equation of state, ${w}_{effT}=-1+\sum_{j}{\ga_{j}\Omega_{j}}$, does not cross the phantom line neither the effective dark energy equation of state,  ${w}_{effT}=-[\alpha\rho'+\rho_{x}]/\rho_{x}$. Their values at $z=0$ vary over the following interval for ${w}_{effT0}\in [-0.749, -0.705] $ and for ${w}_{effx0}$ is around $-0.99$ [see Fig. (\ref{F22})].

\begin{table}[ht!]
\centering
\scalebox{0.70}{
\begin{tabular}{|l|l|l|l|l|l|}
  \hline
  \multicolumn{6}{|c|}{${\rm Cosmological}$ ${\rm parameters}$ ${\rm for}$ ${\rm Q=\alpha\ro'}$} \\
  \hline
  ${\rm N^{o}}$ & $q(z=0)$ & $\omega_{effx}(z=0)$ &  $\omega_{effT}(z=0)$ & $\Omega_{x}(z\approx1100)$ & $\Omega_{x}(z\approx10^{10})$ \\
  \hline
  1 & -0.61 & -0.9998 & -0.7402 & $4.0\times10^{-5}$ & $1.5\times10^{-11}$ \\
  \hline
  2 & -0.59 & -0.9999 & -0.7428 & $7.3\times10^{-6}$ & $7.7\times10^{-12}$ \\
  \hline
  3 & -0.59 & -0.9997 & -0.7298 & $7.0\times10^{-5}$ & $7.8\times10^{-11}$ \\
  \hline
  4 & -0.59 & -0.9990 & -0.7301 & $8.1\times10^{-6}$ & $2.6\times10^{-11}$ \\
  \hline
  5 & -0.62 & -0.9999 & -0.7490 & $8.8\times10^{-6}$ & $2.0\times10^{-10}$ \\
  \hline
  6 & -0.56 & -0.9998 & -0.7047 & $6.1\times10^{-6}$ & $3.5\times10^{-12}$ \\
  \hline
\end{tabular}}
\caption{\label{II2} \scriptsize{We show the cosmological parameters derived from the best fits value of 2-D C.L. obtained in Table (\ref{I1}) by varying two cosmological parameters.}}
\end{table}

\begin{figure}[hbt!]
\begin{minipage}{1\linewidth}
\resizebox{1.6in}{!}{\includegraphics{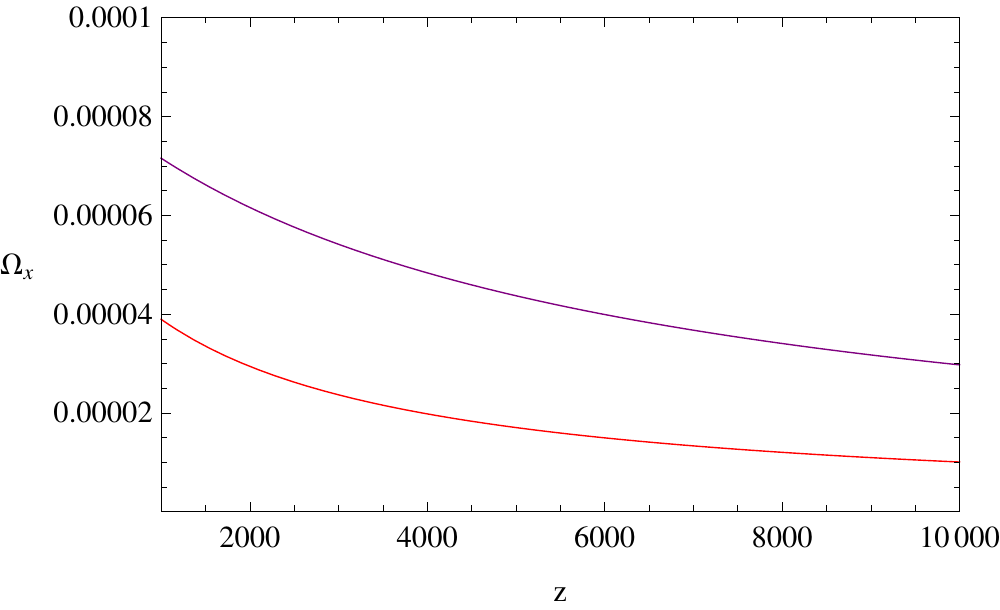}}\hskip0.06cm
\resizebox{1.6in}{!}{\includegraphics{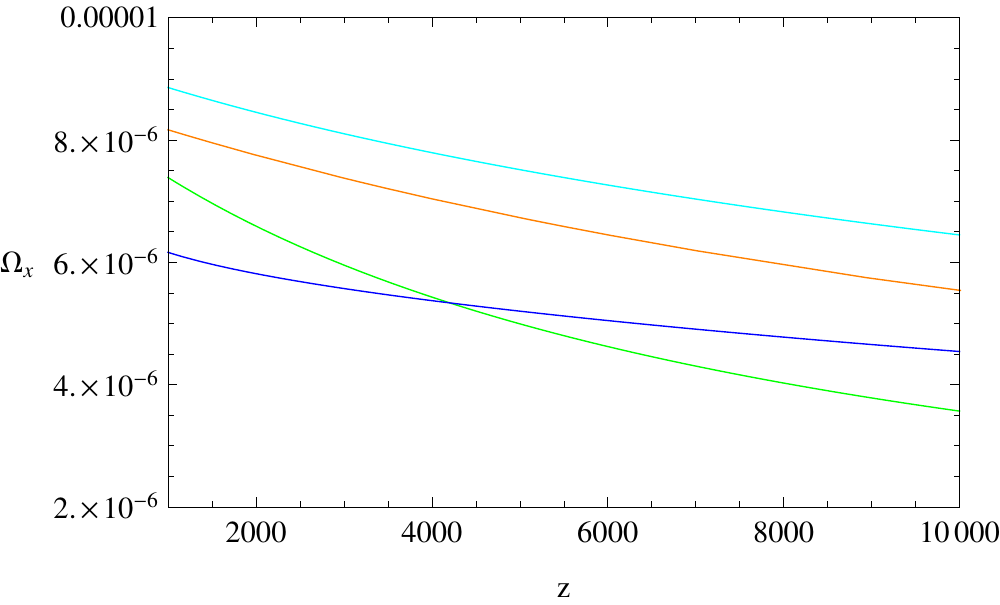}}\hskip0.06cm
\caption{\scriptsize{Plot of $\Omega_{x}(z)$ for $z$ $\epsilon$ $[10^3,10^4]$ using the best-fit values obtained with the Hubble data for different $\theta$ planes. On the left plot, the colors red and purple correspond to the case $1$ and $3$ respectively. In the right plot, the case $2$ is in green color, the case $4$ in orange, case $5$ in cyan and case $6$ in blue.}}
\label{F33}
\end{minipage}
\end{figure}
\vskip 0.2cm

In this model, we look at the behavior of density parameters $\Omega_x$, $\Omega_m$, and $\Omega_r$, so nearly close to $z=0$, see Fig. (\ref{F22}), the dark energy is in particular the main source responsible of the Universe acceleration; far away from $z=1$ the Universe is dominated by the dark matter and at very early times the radiation component enter in action, controlling the entire dynamic of the Universe around $z\simeq10^3$ [cf. Fig.(\ref{F33})]. As it was expected the fraction of radiation at the present moment is negligible; thus, its value varies over the range $10^{-6}\leq\Omega_{r0}\leq10^{-4}$.

At the present, we seek for another kind of constraint that comes form the physics behind recombination or big-bang nucleosynthesis epochs \cite{Luis4}-\cite{Luis7}, this can be considered as a complementary tool for testing our model. As is well known, the fraction of dark energy at the recombination epoch should fulfill the severe bound $\Omega_{x}(z\simeq1100)<0.01$ \cite{Doran}, for the consistence of the model with the big bang nucleosynthesis (BBN) data. Some light could come from the early dark energy (EDE) models, uncovering the nature of DE, at high redshift, as well as their properties, giving an invaluable guide to the physics behind the recent speed up of the Universe \cite{Cala}. The latest constraints on EDE come from the Planck+WP+high L data: $\Omega_{ede} < 0.009$  at $95\%$ C.L \cite{Planck2013} and in the future, the CMB measurements will put further constraints on EDE. We  found that $\Omega_{x}(z\simeq 10^{3})$ is over the interval $[10^{-6}, 10^{-5}]$, so our estimations satisfied the bound reported by the Planck mission [see Table (\ref{II2})]. In addition,  WMAP7+SPT+BAO+SNe leads to $\Omega_{ede} < 0.014$, and WMAP+SPT gives $\Omega_{ede} < 0.013$ \cite{Hou}. Our value on $\Omega_{x}(z\simeq 1100)\leq 10^{-6}$ at the $1\sigma$ level is below the bounds achieved with the forecasting method applied to the Euclid project \cite{HollEuc}; this study is expected to constrain as $\Omega_{ede} < 0.024$. Furthermore, we fulfill the bound reported from the joint analysis based on  Euclid+CMBPol data, $\Omega_{ede} <0.00092$ [see Table (\ref{II2})]. Our estimation on $\Omega_{x}(z\simeq 1100)$ is smaller than the bounds obtained by means of the standard Fisher matrix approach applied to the Euclid and CMBPol experiments \cite{HollEuc}, \cite{Cala1}. In the nucleosynthesis epochs, around $z=10^{10}$, we have that $\Omega_{x}\in [10^{-12}; 10^{-10}]$ at the $1\sigma$level, so the model is in concordance with the conventional BBN processes that occurred at a temperature of $1 {\rm Mev}$ \cite{WrightBBN}.

As is well known, in the early universe, the most important contribution is given by the radiation energy density, that behaves as $\ro\propto(1+z)^{4}$, whereas the contribution of the dark sector is negligible. On the other hand, for a distant future $(z<0)$ the dominant contribution will be composed by the dark energy density while that associated to other beams could be ignored. Under the last limit, the dark energy density of the model behaves as $\ro_{x}(z)=\Lambda\approx\alpha H^{2}(z)+C$. So, when $z\rightarrow-1$ the model goes to a de Sitter stage, an effective cosmological constant dominated era, $H\approx H_{0}\sqrt{\Omega_{x0}+(1-1/(1-\alpha))\Omega_{m0}}$. The constrain made for the interaction $Q=\alpha\ro'$, yields a constant interaction $\alpha = 0.00007^{+0.06604}_{-0.06000}$ that overlaps the one found in \cite{BasilakosI}. Still, it is worth mentioning that this result would indicate that the discussed interaction is highly unlikely. However, the advantage of this model is that we only propose the form of the interaction between two fluids, and this allows us to describe the basic features of the early and the late cosmos.

\section{Cosmic age problem}
With the hypothesis that the universe cannot be younger than its constituents (see \cite{Alcaniz}), we turn our attention to the age problem. The age problem becomes serious when we consider the age of the universe at high redshift. There are some old high redshift objects (OHROs) discovered, for example, the $3.5$ Gyr old galaxy LBDS 53W091 at the redshift $z=1.55$ \cite{Dunlop1}, \cite{Spinrad}, the $4.0$ Gyr old galaxy LBDS 53W069 at $z = 1.43$ \cite{Dunlop2}, the $4.0$ Gyr old radio galaxy 3C 65 at $z = 1.175$ \cite{Stockton}, and the high redshift quasar B1422+231 at $z = 3.62$ whose best-fit age is $1.5$ Gyr with a lower bound of $1.3$ Gyr \cite{Yoshii}. Moreover the old quasar APM 08279+5255 at $z = 3.91$, whose age is estimated between $2.0 - 3.0$ Gyr \cite{Hasinger}, \cite{Komossa}, is used extensively. For this cosmic age, we follow Ref. \cite{Wei} and use the lower bound estimated $2.0$ Gyr at z = 3.91. To give more robustness to our analysis, we use the $0.62$ Gyr gamma-ray burst GRB 090423, at the redshift $z = 8.2$ \cite{Tanvir}, \cite{Salvaterra} detected by the Burst Alert Telescope (BAT) on the Swift satellite in 2009. There are some works that have examined the cosmic age problem within the framework of the dark energy models, see e.g. \cite{Alcaniz}, \cite{Tong}-\cite{Luis3} and references therein. In this section, we would like to consider the age problem for an interaction in the dark sector model.

Given a cosmological model, the cosmic age of our universe at redshift $z$ can be obtained from the dimensionless age parameter
\be
\label{T}
T_{z}(z)=H_{0}t(z)=H_{0}\int^{\infty}_{z}\frac{dz^{\prime}}{(z^{\prime}+1)H(z^{\prime})}.
\ee
At any redshift, the cosmic age of the universe should be larger or equal than the age of the old high redshift objects
\be
\label{S}
T_{z}(z)\geq T_{obj}=H_{0}t_{obj},~ or ~ S(z)=\frac{T_{z}(z)}{T_{obj}}\geq1,
\ee
where $t_{obj}$ is the age of the OHRO. It is worth noting that from Eq. (\ref{T}), $T_{z}(z)$ is independent of the Hubble constant $H_{0}$. On the contrary, from Eq. (\ref{S}), $T_{obj}$ is proportional to $H_{0}$ that we consider as the $H_{0}$ of each case analyzed. In Table \ref{TableSz}, we show the ratio $S(z) = T_{z}(z)/T_{obj}$ at $z = 1.175, 1.43, 1.55, 3.62, 3.91, 8.2$ taking into account the best-fit values obtained in the last section. We find that $T_{z}(z) > T_{obj}(z)$ at  $z = 1.175, 1.43, 1.55, 3.62, 8.2$ but $T_{z}(z) < T_{obj}(z)$ at $z=3.91$, so the old quasar APM 08279+5255 cannot be accommodated as the others old objects. We make this for all cases (see Table \ref{TableSz}). As it is obtained for the $\Lambda$CDM or other models \cite{Wei}, \cite{Cui}, \cite{Forte}, the age problem could be alleviated by taking into account another interaction, for instance a non-linear interaction between the dark components. This fact will be explored in a future work.
\begin{table}[ht!]
\centering
\scalebox{0.85}{
\begin{tabular}{|l|l|l|l|l|l|l|l|}
  \hline
  $N°$ & $S(1.175)$ & $S(1.43)$ & $S(1.55)$ & $S(3.62)$ & $S(3.91)$ & $S(8.2)$ & $t_0$ \\
  \hline
  1 & 1.3425 & 1.1486 & 1.2255 & 1.1928 & 0.8167 & 1.0237 & $13.840^{+0.121}_{-0.070}$ \\
  \hline
  2 & 1.3462 & 1.1519 & 1.2291 & 1.1974 & 0.8199 & 1.0294 & $13.852^{+1.006}_{-0.905}$ \\
  \hline
  3 & 1.3356 & 1.1408 & 1.2164 & 1.1757 & 0.8045 & 1.0029 & $13.948\pm2.180$ \\
  \hline
  4 & 1.3355 & 1.1408 & 1.2163 & 1.1757 & 0.8045 & 1.0033 & $13.949\pm2.671$ \\
  \hline
  5 & 1.3107 & 1.1147 & 1.1861 & 1.1153 & 0.7609 & 0.9201 & $13.911\pm1.110$ \\
  \hline
  6 & 1.3776 & 1.1854 & 1.2681 & 1.2809 & 0.8806 & 1.1514 & $13.944\pm0.412$ \\
  \hline
\end{tabular}}
\caption{\label{TableSz} \scriptsize{It shows the ratio $S(z)=T_{z}(z)/T_{obj}$ at $z = 1.175, 1.43, 1.55, 3.62, 3.91, 8.2$ and the age of the universe today (in Gyr) for the six analyzed cases.}}
\end{table}

\begin{figure}[ht]
\begin{center}
\includegraphics[width=8cm]{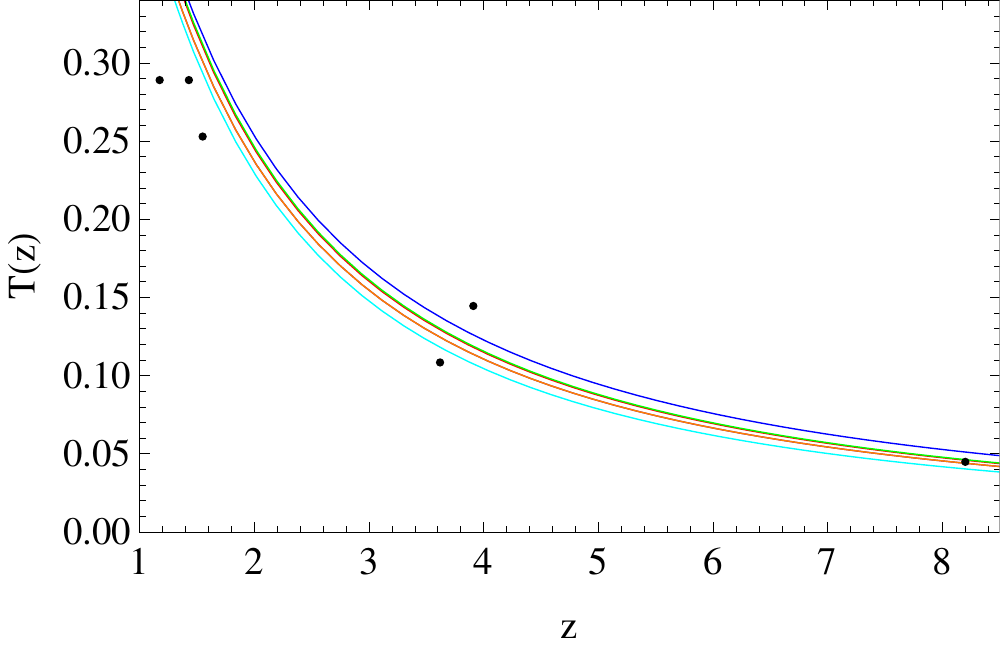}
\caption{\scriptsize{The cosmic age $T$ as a function of the redshift $z$. The graphic shows the six cosmic age curves for the cases $1-6$. The black dots represent the dimensionless age of the OHROs, under the assumption of $H_{0}=70.63$ ${\rm kms^{-1}Mpc^{-1}}$, that correspond to the case $1$.}}
\label{CosmicAge}
\end{center}
\end{figure}

In Table \ref{TableSz}, we see that the value of $S$ at $z=3.91$ is around $0.8$, for all cases, which is far from solving the cosmic age problem. In addition, Fig. (\ref{CosmicAge}) shows the $T(z)$ curves for all cases analyzed. Where for case $1$ corresponds the red color curve, for case $2$ the green, case $3$ purple, case $4$ orange, case $5$ cyan and for case $6$ blue, and the black dots correspond to the first case under the assumption of $H_{0}=70.63$ ${\rm kms^{-1}Mpc^{-1}}$.

Additionally, we have calculated the age of the universe today $t_0$ in Gyr units, the values obtained $t_{0} \in [13.840^{+0.121}_{-0.070}; 13.949\pm2.671]{\rm Gyr}$ are in concordance with the current estimated data released by Planck Mission \cite{Planck2013} or with the data coming from  the WMAP-9 project \cite{WMAP9} [see table (\ref{TableSz})]. In fact, Planck+WMAP data indicate that the actual cosmic age is $13.817\pm0.048$ at $68\%$ C.L,  Planck+WMAP+high L data lead to  $13.813\pm0.047$ at $68\%$ C.L whereas the joint statistical analysis on  Planck+WMAP+high L+BAO  gives $13.798\pm 0.037$ at $1\sigma$ level \cite{Planck2013}. A low value of $t_{0}$ has been found in the recent WMAP-9 analysis, $13.75\pm0.12$, \cite{WMAP9}.

\section{Summary}
We have investigated a Universe that has an interacting dark sector, and two decoupled components, one that could mimic a radiation term and the other which is a baryon component. We have constrained the cosmic set of parameters by using the updated Hubble data and the severe bounds for dark energy found at early times. We have introduced a linear interaction between the dark matter and the dark energy densities in the derivative of the energy density of the dark sector $Q=\alpha\ro'$, and solved the source equation for the total energy density. The model interpolates between a radiation era at early times and a de Sitter phase in the far future, going through a cold dark matter regime.

On the observational side, in the case of 2D C.L., we have made six statistical constrains with the updated Hubble data [see Fig. (\ref{F22}) and Table (\ref{I1})]. Using the priors ($H_{0}$, $\Omega_{b}$, $\alpha$, $\ga_{r}$, $\ga_{m}$)$=$($69$, $0.0449$, $0.0001$, $4/3$, $1.014$), the best-fit values for the present-day density parameters are given by  $(\Omega_{x0},\Omega_{m0})=(0.733^{+0.068}_{-0.072}$, $0.222^{+0.108}_{-0.107})$; so the results obtained are in agreement with the data released by Planck Mission \cite{Planck2013} or with the data coming from  the WMAP-9 project \cite{WMAP9} [see Fig. (\ref{FO})]. Besides, it turned out that $H_{0} \in [69.92^{+2.11}_{-2.15}; 70.77^{+2.29}_{-2.27} ]{\rm km~s^{-1}\,Mpc^{-1}}$ so the latter values are met within $1\sigma$ C.L. reported by the Planck+WMAP+high L analysis \cite{Planck2013} and from the recent WMAP-9 analysis \cite{WMAP9}. One of the best-fit values of the barotropic index of the dark mater, $\ga_{m}=1.049^{+0.181}_{-0.460}$, has a small difference with dust index, equal to one, so the dark matter is not pressureless. Moreover, the deceleration parameter at the present time $q(z=0)\in[-0.62,-0.56]$ in agreement with \cite{WMAP9} and the total equation of state today, $w_{effT}(z=0)\in[-0.749,-0.705]$; indeed, $-1\leq w_{effT}\leq0$, and therefore does not cross the phantom barrer [see Fig. (\ref{F22})], while the fraction of radiation at the present moment $\Omega_{r0}$ varies from $2\times 10^{-6}$ to $0.0001$ for the six cases mentioned in Table (\ref{I1}). 

We have constrained the behavior of dark energy in the recombination era and compared it with the latest bounds coming from the Planck+WP+high L data, SPT, and ACT, among other experiments. We have found that $\Omega_{x}(z\simeq 10^{3}) \in [10^{-6}, 10^{-5}]$, therefore our estimations satisfied the stringent bound reported by the Planck mission, $\Omega_{ede} < 0.009$  at $95\%$ C.L. \cite{Planck2013} [see Table (\ref{II2})] and agreed with  the small-scale CMB temperature measurement from the SPT \cite{Reic} or with the upper limit set by WMAP7+SPT+BAO+SNe data \cite{Hou}. Moreover, the value $\Omega_{x}(z\simeq 10^{3})$ obtained in this work will be consistent with the future constraints achievable by the Euclid and CMBPol experiments, \cite{HollEuc}, \cite{Cala1}. Around $z=10^{10}$, the nucleosynthesis epoch, the dark energy found fulfills the strong upper limit $\Omega_{x}(z\simeq 10^{10})<0.04$ at the $68\%$ C.L. \cite{WrightBBN}, so the standard BBN processes and the well-measured abundance of light elements are not disturbed.

In addition, we have obtained that the age of the universe $t_{0} \in [13.840^{+0.121}_{-0.070}, 13.949\pm2.671]$ Gyr is in agreement to the one reported by WMAP-9 year project \cite{WMAP9} and the one reported by the Planck Mission, \cite{Planck2013}.

\acknowledgments
The author would like to thank Prof. L. P. Chimento for useful comments that greatly improved the clarity of the manuscript. Also acknowledges the support of CONICET, IMAS and Math. Department, FyCEN-UBA.



\vskip 1cm


\end{document}